\documentclass[twocolumn,aip,reprint]{revtex4-1}
\usepackage{lineno}

\usepackage[percent]{overpic}
\usepackage{etoolbox}

\usepackage{graphicx,framed,amssymb,mathrsfs,textcomp,comment,fancyhdr,calligra,ifsym,amsmath,amsthm,mathtools,amsmath,float}

\usepackage[T1]{fontenc}
\usepackage[utf8]{inputenc}
\begin{document}
\title{{\bfseries A hybrid ring design in the storage-ring proton electric dipole moment experiment}}
\author{S. Hac{\i}\"{o}mero\u{g}lu$^{1}$, Y.K. Semertzidis$^{1,2,*}$}
\affiliation{
$^{1}$Center for Axion and Precision Physics Research, Institute for Basic Science (IBS/CAPP), Daejeon 34051, Republic of Korea\\
$^{2}$Department of Physics, Korea Advanced Institute of Science and Technology (KAIST), Daejeon 34141, Republic of Korea\\
$^{*1}$Corresponding author, yannis@kaist.ac.kr
}

\renewcommand{\[}{\begin{equation}}
\renewcommand{\]}{\end{equation}}
\renewcommand{\[}{\begin{equation}}
\renewcommand{\]}{\end{equation}}
\renewcommand{\v}{{\bf v}}
\newcommand{\E}{{\bf E}}
\newcommand{\B}{{\bf B}}
\newcommand{\s}{{\bf s}}
\renewcommand{\b}{\boldsymbol\beta}
\renewcommand{\d}{{\bf d}}
\newcommand{\m}{{\boldsymbol\mu}}
\newcommand{\om}{\boldsymbol\omega}
\date{\today}
\begin{abstract}

A new, hybrid design is proposed to eliminate the main systematic errors in the frozen spin, storage ring measurement of the proton electric dipole moment.  In this design, electric bending plates steer the particles, and magnetic focusing replaces electric. The magnetic focusing should permit simultaneous clock-wise and counter-clock-wise storage to cancel systematic errors related to the out-of-plane dipole electric field. Errors related to the quadrupole electric fields can be eliminated by successive runs of magnetic focusing with different strengths.

\end{abstract}
\maketitle

\section{Introduction}

CP-violation is a necessary condition of the observed baryon-antibaryon asymmetry in our universe.~\cite{sakharov} However, the CP-violating phase in weak interactions and the P- and T-violating {$\theta_{\rm QCD}$-term}  of quantum chromodynamics (QCD) in the standard model (SM) are not enough to explain that asymmetry. A new, much stronger CP-violating source is needed, probably from physics beyond the SM, e.g., super-symmetry (SUSY). EDM searches are some of the most sensitive probes of physics beyond the SM associated with CP-violation.~\cite{landau,susyref,lrsref,higgsref,dipole_moments,engel13}

The neutron EDM (nEDM), electron EDM (eEDM), and $^{199}$Hg EDM experimental sensitivities are currently the most advanced ones and set the most restrictions in the relevant parameter space.  The current (direct) limit of the nEDM is $<3 \times 10^{-26}\, {\it e} \cdot {\rm cm}$,~\cite{baker} the eEDM limit (indirect, from the study of ThO molecules) is $<9 \times 10^{-29}\, {\it e} \cdot {\rm cm}$,~\cite{baron} and the $^{199}$Hg experimental EDM limit $<8 \times 10^{-30}\, {\it e} \cdot {\rm cm}$.~\cite{moledm}
The SM contribution from the electro-weak sector for hadronic EDMs is at the $10^{-30}-10^{-31}\, {\it e} \cdot {\rm cm}$ level, while the {$\theta_{\rm QCD}$-term}  contributes~\cite{pospelov} as

\begin{equation}
d_n = \theta_{\rm QCD} \times 3.6 \times 10^{-16}\, {\it e} \cdot {\rm cm}
\end{equation}
meaning from the nEDM limit that $ \theta_{\rm QCD} < 10^{-10}$, while from the theory of QCD it was expected to be of order 1.  To solve this apparent discrepancy, for an otherwise very successful theory, Peccei and Quinn came up with a new symmetry,~\cite{peccei} the breakdown of which requires the existence of axions.~\cite{weinberg,wilczek,kim,shifman,zhitnitsky,dine}  With or without this axion field, the contribution of the {$\theta_{\rm QCD}$-term} can be anywhere below the current experimental limits.  In addition, physics beyond the SM, e.g., SUSY, is also possible anywhere below the present experimental limits.  The contributions of the different potential sources combine in different ways in different systems, so more than one system is needed to decipher the CP-violating source should one be found to be non-zero.~\cite{engel13}  The mercury atomic EDM limit translates to $2 \times 10^{-26} \, {\it e} \cdot {\rm cm}$ for the neutron and $5 \times 10^{-25} \, {\it e} \cdot {\rm cm}$ for the proton, i.e., setting the most precise experimental limits for those systems.

The storage ring EDM method is proposed to reach below $ 10^{-29} \, {\it e} \cdot {\rm cm}$, targeting  $ 10^{-30} \, {\it e} \cdot {\rm cm}$ with an upgrade.  This sensitivity goal is possible due to the high intensity polarized beams available, the long spin coherence time, and the high analyzing power for magic momentum protons.  In this document we will suggest a modification to the original all-electric storage ring method to reduce the risk of the main systematic error, that of the radial B-field present around the ring. The new method uses a series of different strength measurements using magnetic alternating focusing in order to be able to differentiate between a genuine EDM effect and an effect originating from other, background-related sources.  According to our precision simulations we expect to be able to simultaneously store counter-rotating beams, which will help  minimize systematic errors by several orders of magnitude.  Some runs may include different strength electric focusing for systematic error studies.

\section{Experimental Method using a hybrid ring lattice}

The magnetic dipole moment (MDM) $\m$, and EDM $\d$ of a particle with rest-frame spin $\s$, charge $e$, and mass $m$ are defined as $\m =(ge/2m)\s$ and $\d = (\eta e/ 2mc)\s$, with $g$ the g-factor and $\eta$ playing the same role for EDM as $g$ is playing for the MDM.  At rest, the spin precession rate of a particle in magnetic and electric fields is given by
\[\frac{d\s}{dt} = \m\times\B + \d\times \E.\]
However, in a storage ring the spin precession is modified by the Thomas term due to the angular acceleration and it is described by the T-BMT~\cite{bmt,fukuyama,khriplovich} equation, assuming $\b \cdot \B = \b \cdot \E = 0$,
\[\begin{aligned}\om_a& = \frac{e}{m}\left[ G\B  -\left(G - \frac{1}{\gamma^2 - 1}\right)\frac{\b\times\E}{c}\right]  \\\om_e&=  \frac{\eta e}{2 m}\left(\frac{\E}{c} + \b\times\B \right),\end{aligned}\]
where $G = (g-2)/2$, $\beta=v/c$, and $\gamma$ is the relativistic Lorentz factor.  It is evident that for a charged particle at rest, applying a net electric field to probe its EDM is not an option, as the particle will be accelerated and get lost very fast.  Even though the EDM of charged particles has been probed in storage rings,~\cite{muonedm1,muonedm2} one way to improve the sensitivity to the EDM of charged particles is the frozen spin method,~\cite{far,orlov,nel,mane1,mane2,protonEDMprop,rsi} where the  spin is kept along the momentum vector and the radial electric field is acting on the particle EDM for the duration of the storage time.  An all-electric ring, albeit smaller by an order of magnitude, is already constructed and operated at Heidelberg.~\cite{essr3}
The all-electric ring allows for simultaneous clock-wise and counter-clock-wise storage.  A net radial B-field around the ring ($N=0$, with $N$ the Fourier component of the radial B-field around the ring azimuth) will cause an EDM-like signal by precessing the particle spin similarly to an EDM precession.  

The new experimental approach suggested here is superior to the previously proposed method  in that by using a series of measurements with different strength alternating magnetic focusing  we can distinguish a genuine EDM signal from a background one, reducing the effect of major potential systematic error sources by several orders of magnitude.  As part of the systematic error studies, weak electric focusing may also be included.  The simulations used here are based on Runge-Kutta integration.~\cite{hamming59,selcuk,mane15}

The radial B-field modes constitute a serious systematic error in the all-electric storage ring EDM experiment.  Our plan to reduce this field with shielding works best when the vertical beta-function is uniform around the ring.  In that case, only the $N=0$ mode of the radial B-field produces a non-zero vertical spin precession rate.  The effect of the non-uniform vertical beta-function~\cite{carli} is to produce a significant vertical spin precession rate due to the high $N$ modes of the radial B-field.   Here we summarize the current situation and suggest a way to differentiate the effect of the radial B-field from a genuine EDM effect after applying magnetic focusing instead of electric focusing.
The original idea to cancel the radial B-field effect was based on the assumption that the main background originates {\it only}
 from the $N=0$ (DC) component of the radial B-field.  The higher modes were considered to be harmless, except from the possibility of confusing the interpretation of the SQUID-based beam position monitors (BPMs).  Hence,

\begin{enumerate}
\item Our plan was to reduce the BPM sensitivity to higher modes. Shield the magnetic fields to 1-10\,nT everywhere around the ring.  Even though the geometrical phase effects due to the magnetic fields cancel with CW and CCW injections, this level of shielding is considered easily achievable with present day technology, so it is our goal.~\cite{selcuk_sh}
\item The time stability of the shielded horizontal B-field from our measurements is at the level of 100\,pT per 10 minutes with a longitudinal gradient of its radial component of about 5\,pT/m.  This time stability is adequate for the needs of the experiment to the required level.~\cite{selcuk_sh}
\item The goal of reducing the higher than $N=0$ radial B-field modes was solely based on the impression that they were harmless regarding the spin precession rate and therefore all the effort was focused on reducing their impact on the SQUID-based BPM output.  Indeed, the SQUID-based BPM signals in a lattice with uniform beta-functions scale as $(Q_y/N)^4$ for time modulated focusing quads (and for $N>1$ value), so that they would not compete as a background with the EDM signal.~\cite{rsi,kawall}
\end{enumerate}

After Christian Carli demonstrated that the non-uniform vertical beta-function has a profound effect on the vertical spin precession rate we considered a number of possible plans to reduce the experimental sensitivity to radial B-field modes with $N>10$.  The value of $N>10$ was chosen because we had assumed 20 SQUID-based BPMs, equally spaced around the ring and hence we could detect all the modes up to and including the $N=10$ mode.~\cite{rsi}

\begin{enumerate}
\item One of us (S. H.) suggested using a combination of different lattices to decipher the beam dynamics resonances responsible for the vertical spin precession rate.  The studies have shown significant promise to warrant more investigation in this direction.
\item Design a ring with as uniform vertical beta-function as possible.  Having a completely uniform focusing ring may be possible; however, 
taking this approach will require a long and detailed study
to make it compatible with all the additional requirements and effects in our experiment.  It will also require a long and detailed study on the subject to come up with a credible lattice design and even though it may be worthwhile to explore this option, next we propose a simpler approach.
\item The new idea is to use magnetic instead of electric focusing in the ring.  We have demonstrated that we could store beams CW and CCW at the same time when using alternating magnetic focusing, making this possibility feasible. This fact is critical since the inescapable vertical E-field effect from a misaligned radial E-field plate is also an important systematic error source.~\cite{deuteron}

\end{enumerate}

\begin{figure}[t]
\centering
\includegraphics[scale= 0.4]{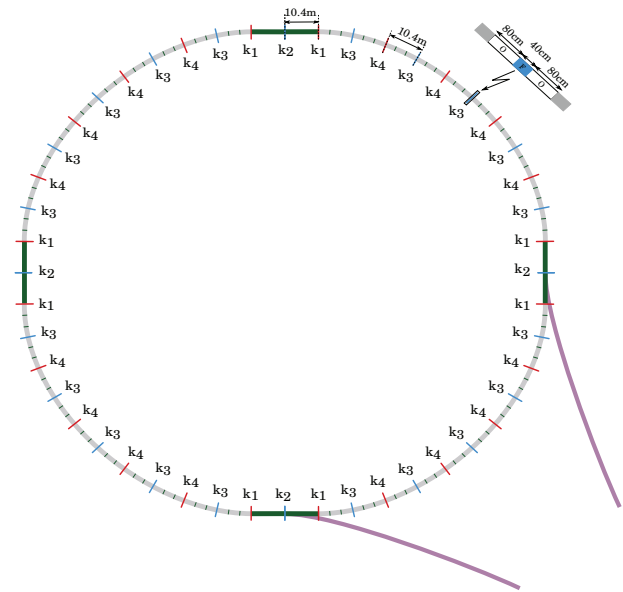}
\vspace*{-4cm}
\caption{\label{lattice2}A detail of the storage ring lattice is shown here with focusing and defocusing quadrupoles (shown as $\rm k_3$ and $\rm k_4$).  The bending sections, including the short straight sections, have a length of 10.417\,m, three sections assembled as one unit.  The long straight sections are 20.834\,m long with a quadrupole (shown as $\rm k_2$) in the middle and two half-length quads (shown as $\rm k_1$) at both ends.  The values of the magnetic quadrupole strength are: $k_1 = 0.1$T/m, $k_2 = -0.1$T/m, $k_3 = -0.1$T/m, $k_4 = 0.1$T/m.  The vertical tune, when running with these quadrupole strengths, is $Q_y = 0.67$, while the horizontal tune is $Q_x = 1.73$.  
}
\end{figure}

\begin{figure}[t]
\centering
\vspace*{0cm}\includegraphics[scale = 1]{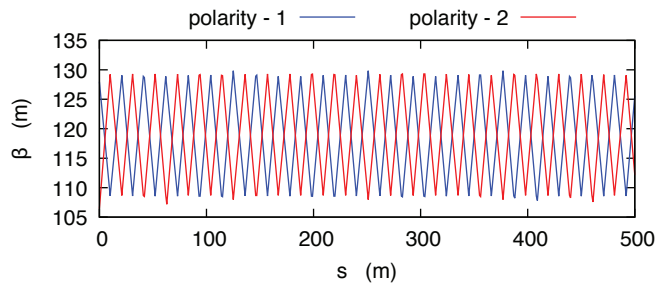}
\vspace*{-0.4cm}\caption{\label{ver_beta_function_fin}The vertical beta-function values around the ring for CW and CCW operations.  They flip sign when the magnetic quadrupoles are running with the opposite sign and therefore the counter-rotating particles on average trace the same paths.
}
\end{figure}

\begin{figure}[t]
\centering
\vspace*{0cm}\includegraphics[scale = 1]{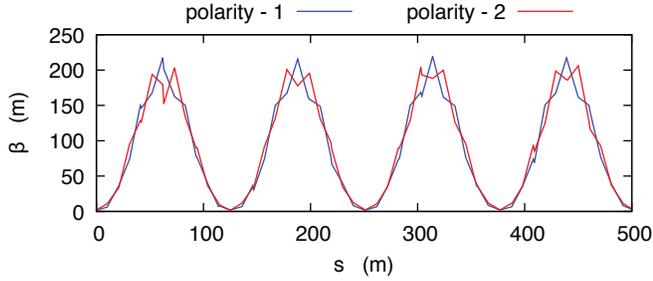}
\vspace*{-0.4cm}\caption{\label{hor_beta_function_fin}The horizontal beta-function values around the ring for CW and CCW operations.
}
\end{figure}

The counter-rotating beams do not actually go through the same places everywhere, due to the fact that the vertical focusing includes magnetic focusing. Therefore, those beams may not exactly cancel those systematic errors at all places.  However, we have shown that it is possible to use the same magnetic quads with flipped field directions (opposite sign currents) and on average the particles do follow the same trajectories.  The study showed that this approach is very promising.  This idea seems to work perfectly well, eliminating completely the radial B-field issue.  In addition, the vertical dipole E-field effect is cancelled completely in CW and CCW injections as is the effect of gravity.  The suggested working lattice is shown in Figure \ref{lattice2}, which is a modification of the lattice shown in the paper~\cite{rsi} describing the all-electric storage ring method, but this time the electric quadrupoles are replaced with corresponding magnetic ones.  Figure \ref{ver_beta_function_fin} shows the vertical beta-function of the CW and CCW stored beams, and Figure \ref{hor_beta_function_fin} the corresponding for the horizontal.  Flipping the sign of the currents in the magnetic quadrupoles will produce symmetric beta-functions for the CW and CCW beams. 

However, it is always possible that some electric focusing will be present somewhere in the ring.  This focusing and/or defocusing could originate from the bending electric field plates, which produce the required radial E-field. One or both plates could be misaligned, readily producing a vertical dipole, but also a quadrupole or even higher multipole E-fields.  There could also exist induced charges (image charges) from any horizontally placed metals around the lattice, the tune shift and tune spread effects due to high beam intensities, etc.  Some of those systematic errors we may be able to detect, e.g., by modulating the voltage on the bending E-field plates or control them by using beam bunch intensities of various strengths.  At the end of the experiment, however, we need to have high confidence regarding the origin of the effect.  Here we are suggesting using a number of runs with different vertical magnetic focusing strengths in order to differentiate between a systematic error and a genuine EDM signal.

\begin{figure}[t]
\centering
\vspace*{0cm}\includegraphics[scale = 1.2]{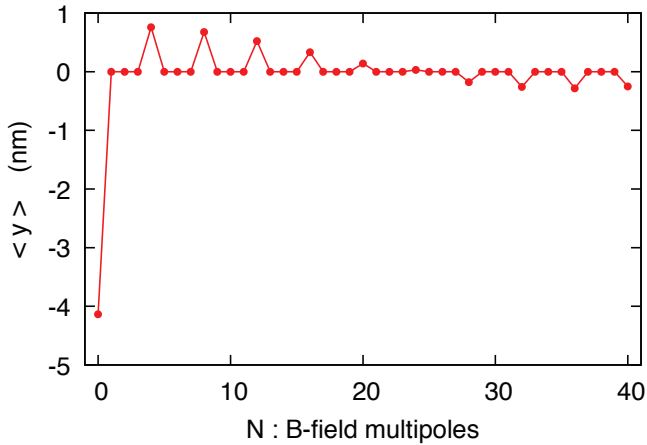}
\vspace*{-0.4cm}\caption{\label{y_vs_n}The average vertical beam offset when only magnetic focusing is used, as a function of the radial B-field multipoles ($N$-values).  The amplitude of the background radial B-field is always kept at 1\,pT, while the quadrupole strength is kept at $\pm$0.1T/m.}
\end{figure}

\begin{figure}[t]
\centering
\vspace*{0cm}\includegraphics[scale = 1.2]{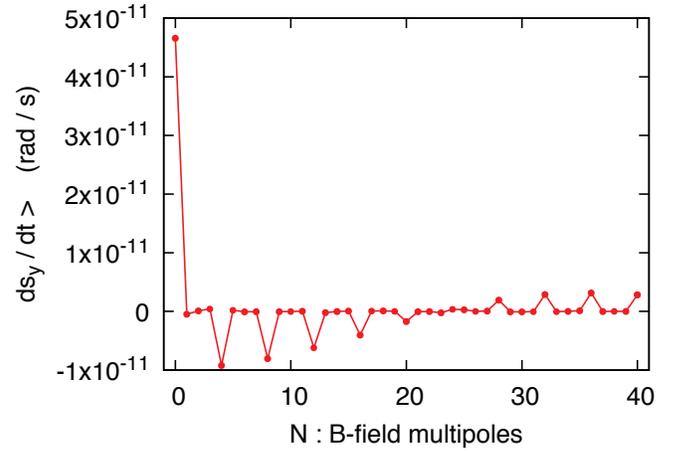}
\vspace*{-0.4cm}\caption{\label{wr_vs_n}The vertical spin precession of the counter-rotating beams when only magnetic focusing is used, for different radial B-field multipoles ($N$-values).  The amplitude of the radial B-field is always kept at 1\,pT, while the quadrupole strength is kept at $\pm$0.1T/m. A genuine EDM signal for $ 10^{-29} \, {\it e} \cdot {\rm cm}$ is larger than 1\,nrad/s, and therefore much larger than the background signal.}
\end{figure}

\begin{figure}[t]
\centering
\vspace*{0cm}\includegraphics[scale = 0.4]{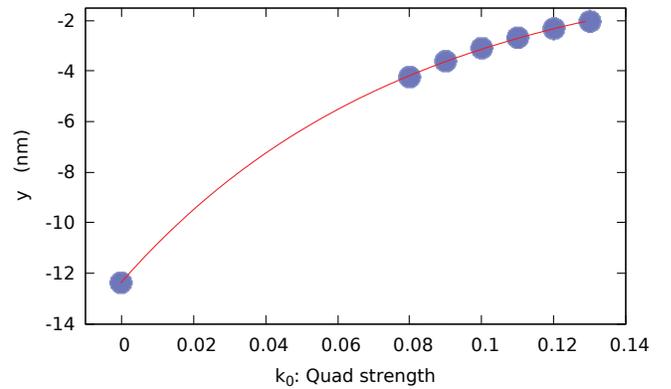}
\vspace*{-.0cm}\caption{\label{y_vs_k}The average vertical beam offset vs. the magnetic quadrupole strength in units of T/m with alternating sign quadrupoles. In addition to the magnetic focusing there is also electric focusing with a field index of $m=Q_y^2=0.1$, with $Q_y$ the vertical tune when there is only electric focusing.  The radial B-field applied is $B_r = $1pT, and $N=0$ for its harmonic value around the ring. The vertical offset does not go to zero for zero magnetic focusing due to the presence of the electric focusing.}
\end{figure}

\begin{figure}[t]
\centering
\vspace*{0cm}\includegraphics[scale = 0.85]{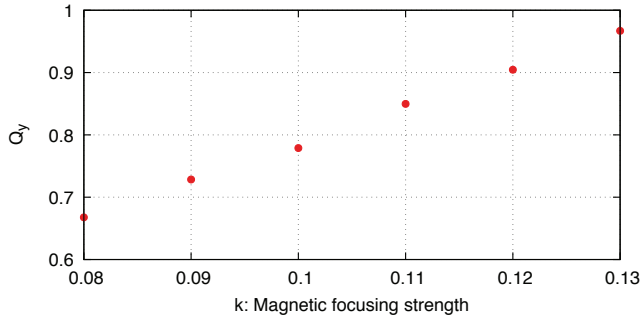}
\vspace*{-.3cm}\caption{\label{Qy_vs_k}The vertical tune vs. the magnetic quadrupole strength in units of T/m with alternating sign quadrupoles. In addition to the magnetic tune, there is also electric focusing due to the special shaping of the electric field plates with a field focusing index of $m=0.1$.
}
\end{figure}

The total effect, i.e. the vertical spin precession rate, is going to be in a functional form: 
\[R_V = R_{\rm EDM} + R_{\rm B_r} \times \frac{Q_{\rm Backgr}^2  + ...} {\zeta \times Q_{\rm Magnetic}^2 +Q_{\rm Backgr}^2 + ...}\]
where $R_V$ is referring to the total vertical spin precession rate, $R_{\rm EDM}$ refers to the portion due to the particle EDM, 
$Q_{\rm backgr}^2 =f( Q_{\rm Electric}^2, Q_{\rm ImageCharge}^2, Q_{\rm BeamIntensity}^2,...)$ corresponds to the square of the tuning due to non-magnetic effects, $Q_{\rm Magnetic}^2$ is the square of the tune due to the magnetic quads, $Q_{\rm Electric}^2$, $Q_{\rm ImageCharge}^2$, $Q_{\rm BeamIntensity}^2$ are the square of the tunes due to the electric quads, the forces due to induced charges, and the forces due to the beam intensity, correspondingly.  $R_{Br}$ refers to the vertical spin precession rate due to the radial B-field.  The point is that a net radial B-field can create a vertical spin precession, which can only be canceled exactly by another B-field; in this case we assumed it to be the magnetic focusing.  Magnetic focusing can essentially eliminate this systematic error provided that it is the only source focusing the beam.    Figure \ref{y_vs_n} shows the average vertical offset of the stored beam as a function of the radial B-field multipole whose amplitude is always kept at 1\,pT.  Figure \ref{wr_vs_n} shows the vertical spin precession rate under the same conditions.  A genuine EDM signal for $ 10^{-29} \, {\it e} \cdot {\rm cm}$ is larger than 1\,nrad/s, and therefore much larger than the above background signal.  However, if on one of the magnetic quadrupoles we add an overlapping electrical quadrupole with a strength of 1\,kV/m$^2$, then we get the much larger spin precession rate of 0.4\,nrad/s, for $N=4$ harmonic case of the radial B-field.  This effect will be further and effectively suppressed by applying varying levels of magnetic field focusing, as described in the section below.
 Figure \ref{y_vs_k} shows the average vertical offset of the beam as a function of the magnetic focusing strength for the radial B-field $N=0$ multipole whose amplitude is always kept at 1\,pT.   Figure \ref{Qy_vs_k} shows the vertical tune vs. the magnetic focusing strength in the presence of an electrical focusing field of $m=0.1$ due to the shape of the electrical deflectors.   

\begin{figure}[t]
\centering
\vspace*{0cm}\includegraphics[scale = 0.85]{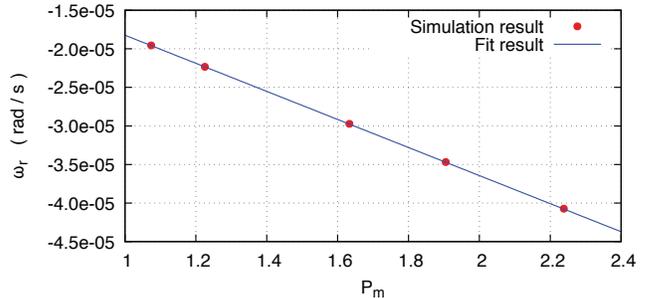}
\vspace*{-0.8cm}\caption{\label{wr_vs_Pm}The vertical spin precession rate as a function of the $P_m = 1/Q_y^2$ when the background effect is due to a combination of a DC ($N=0$) radial magnetic field around the ring with strength of 1\,pT and a large electric focusing effect of the bending plates. The bending plate focusing corresponds to an (electric) vertical focusing field index of $m=0.1$.   The fit result is from a first order polynomial.  The DC-offset corresponds to the EDM precession rate, which in this case is $-1.9 \times 10^{-8}$rad/s, consistent within the estimated errors to the input EDM value corresponding to $-4.1 \times 10^{-8}$rad/s.}
\end{figure}

{\it Experimental Approach.} A practical way to proceed would be to first apply weak E-focusing and run the experiment for a number of fills. We can then apply the appropriate radial B-field to cancel the $N=0$ and higher radial B-field modes depending on the vertical spin precession rates and the SQUID-based BPM signals. (Note: Another large potential systematic error is due to orbit corrugation, i.e., the orbit non-planarity.  However, this effect goes as the sine of the horizontal spin angle relative to momentum and it can be probed by a beam bunch whose spins are frozen in the radial direction.~\cite{deuteron,sredm_coll})  Next, we apply a series of B-field focusing strengths, from weak to stronger ones to probe the EDM effect.  With magnetic focusing the main systematic error is the out-of-plane dipole electric field, which is cancelled by CW and CCW beam storage as in the deuteron storage ring EDM experiment.~\cite{deuteron}  Since simultaneous CW and CCW storage is possible in the current configuration, then most of the issues related to E-field direction stability go away.  
In addition, any focusing effect from the electric field plates or any other sources is sorted out by running the experiment at different alternating magnetic focusing strengths as shown in Figure \ref{wr_vs_Pm}.  Here, an additional electric focusing exists together with a DC ($N=0$) radial magnetic field around the ring with strength of 1\,pT.  The electric focusing is originated by shaping all the bending plates, producing a vertical focusing with a field index of $m=0.1$.  The spin precession rate equation, when expanded, can be written as
\[R_V = R_{\rm EDM} + R_{\rm B_r} Q_{\rm Backgr}^2  P_{m1} - R_{\rm B_r} Q_{\rm Backgr}^4  P_{m1}^2 +  ... \]
with $P_{m1} = 1/(\zeta \times Q_{\rm Magnetic}^2)$, showing clearly that for a large magnetic focusing tune, i.e., $P_{m1} \rightarrow 0$, the spin precession rate corresponds to the EDM signal.  Hence, the DC offset in Figure \ref{wr_vs_Pm} corresponds to the EDM signal and the obtained value is consistent with the simulations.  In Figure \ref{wr_vs_Pm},  the spin precession rate corresponds to $ 10^{-28} \, {\it e} \cdot {\rm cm}$ EDM level to prove the principle of the method.  It will be advantageous to keep the spin precession rate lower by adding much stronger magnetic focusing cases and keep the electric focusing below the $m=0.01$ level.  The method will work best, requiring less leverage, when the magnetic focusing is dominating all other focusing effects.  In a similar way, we can prove that the sextupole vertical electric field cancels with CW and CCW storage, etc., provided that the beam emittances are the same to an adequate level.
From our simulations we infer that the SQUID-based BPMs resolution requirements are relaxed by several orders of magnitude over the lattice where electric focusing is used, which is a major breakthrough.  The new requirements are a well-shaped quadrupole magnetic field in the ring, so that the center of the CW and CCW beams overlap within 100\,nm at all magnetic quadrupole strengths, using the SQUID-based BPM signals.  In addition, the ring needs to be flat (absence of corrugation) to 100\,nrad, which we achieve by a combination of mechanical alignment, beam-based alignment and by using bunches polarized in the radial direction.
 A summary of the main systematic errors in the experiment with hybrid fields (electric bending and magnetic focusing) and their current remediation plan is given in Table \ref{syserr}.

\begingroup
\begin{table}[b]
\caption{\label{syserr}Main systematic errors and their remediation when hybrid fields (electric bending and magnetic focusing) are used.}
\begin{ruledtabular}
\begin{tabular}{ l  | p{5cm} }
{\bf Effect} & {\bf Remediation}  \\ \hline
Radial B-field. &  Magnetic focusing. \\ \hline
Radial B-field when &  Varying magnetic focusing \\
other than magnetic &  and fit for the DC offset in the \\
focusing is present. &  vertical precession rate. \\ \hline
Dipole vertical E-field. &  CW and CCW beam storage. \\ \hline
Corrugated (non-planar) & Probe with stored beams  with their \\
orbit. &  spins frozen in the radial direction.~\cite{deuteron,sredm_coll}\\ \hline
RF cavity misalignment & Vary the longitudinal lattice impedance to probe the effect of the cavity's vertical angular misalignment. CW and CCW beams cancel the effect of a vertically misplaced cavity.~\cite{rsi}
\end{tabular}
\end{ruledtabular}
\end{table}
\endgroup

\section{Conclusions}
We have suggested a hybrid storage ring EDM method, where the focusing is solely done by alternating focusing magnetic fields, while the horizontal bending is accomplished by vertical (ideally strictly cylindrical) electro-static plates.  This configuration allows the simultaneous storage of CW and CCW beams, enabling the cancelation of the main systematic error in this case, that of the dipole vertical E-field.  The effect of the radial B-field as a background is cancelled by the magnetic focusing, while that of the electric quadrupole focusing is
diminished by taking a number of runs with different vertical magnetic focusing strengths.  This new conceptual improvement greatly relaxes the magnetic shielding requirements of the ring, improving the feasibility of the target sensitivity of $10^{-29} \, e\cdot$cm and possibly beyond. 

\section{Acknowledgements}
IBS-Korea (project system code: IBS-R017-D1) supported this project.  We acknowledge useful discussions with members of the srEDM and JEDI collaborations, as well as our colleagues from CERN, specifically Christian Carli, Michael Lamont, Hans Stroeher, Richard Talman and others.  We especially thank Sidney Orlov for greatly improving the readability of the paper.

\end{document}